\documentclass{aa} 

\DeclareRobustCommand{\rchi}{{\mathpalette\irchi\relax}}
\newcommand{\irchi}[2]{\raisebox{\depth}{$#1\chi$}} 

\newcommand{\rlight}{r_{\rm L}}

\newcommand{\rot}{\mathbf{\nabla} \times}
\newcommand{\divg}{\mathbf{\nabla}\cdot}

\usepackage{graphicx}
\usepackage{txfonts}
\usepackage{color}

\begin{document} 

\title{Spheroidal force-free neutron star magnetospheres}


\author{J. P\'etri
}

   \institute{Universit\'e de Strasbourg, CNRS, Observatoire astronomique de Strasbourg, UMR 7550, F-67000 Strasbourg, France.\\
              \email{jerome.petri@astro.unistra.fr}         
             }

   \date{Received ; accepted }

 
\abstract
   {Fast rotating and self-gravitating astrophysical objects suffer strong deformations from centrifugal forces. If moreover they are magnetized, they generate an electromagnetic wave that is perturbed accordingly. When stellar objects are also surrounded by an ideal plasma, a magnetosphere is formed. For neutron stars, a relativistic magnetized wind is launched, efficiently extracting rotational kinetic energy flowing into particle creation and radiation.}
   {We study the electromagnetic configuration of a force-free magnetosphere encompassing an ideal spheroidal rotating conductor as an inner boundary. We put special emphasize to millisecond period neutron star magnetospheres, those showing a significant oblate shape. For completeness, we also investigate the effect of prolate stars.}
   {Force-free solutions are computed by numerical integration of the time-dependent Maxwell equations in spheroidal coordinates using pseudo-spectral techniques. Relevant quantities such as the magnetic field structure, the spin down luminosity, the polar cap rims and the current density are shown and compared to the force-free spherical star results.}
   {We find that the force-free magnetic field produced by spheroidal stars remains very similar to their spherical counterpart. However, the spin down luminosity slightly decreases with increasing oblateness or prolateness. Moreover the polar cap area increases and always mostly encompasses the equivalent spherical star polar cap rims. The polar cap current density is also drastically affected.}
   {Neutron stars are significantly distorted by either rotational effects like millisecond pulsars or by magnetic pressure like magnetars and high-magnetic field pulsars. Observational interpretation and fitting of thermal X-ray pulsation will greatly benefit from an accurate and quantitative analyse like the one presented in this paper. However, even for the fastest possible rotation, the effect would certainly be unobservable, in the sense that we can't predict what feature of the light curve would supposedly reveal the neutron star deformation due to fast rotation. 
   }


\maketitle

%

\section{Introduction}

Self-gravitating astrophysical objects are usually set into rotation due to diverse formation and evolution scenarios. Gravitational attraction is compensated by pressure but also by centrifugal forces. For fast rotating astrophysical objects, the centrifugal force substitutes to a substantial fraction of the internal pressure to sustain the system in a quasi-stationary equilibrium. If moreover, the gas is magnetized, it radiates an electromagnetic wave, removing energy and angular momentum from the system. Single stars or clouds are typical configurations where such phenomena occur.

This problem is particularly relevant in the realm of high energy astrophysics. Indeed, fast rotating neutron stars are privileged places where the system deviates significantly from a spherical shape, like for instance millisecond pulsars. Magnetars are also expected to show strong magnetic constrictions leading to oblate or even prolate surfaces not directly connected to a geometry imposed by their rotation. A detailed analysis of such systems evolving in vacuum was recently performed by \cite{petri_spheroidal_2021}. If vacuum is replaced by an ideal plasma like in electron/positron pairs, a spheroidal force-free neutron star magnetosphere can be constructed, generalizing the extensive literature about the force-free magnetosphere of spherical stars \citep{spitkovsky_time-dependent_2006, komissarov_simulations_2006, mckinney_relativistic_2006, petri_pulsar_2012, cao_spectral_2016}.

From an observational point of view, thermal X-ray emission from neutron star hot spots was used to constrain the compactness and radius of these compact objects \citep{riley_nicer_2019, bogdanov_constraining_2019}. Their estimates rely on fitting the X-ray pulsed profile taking into account general-relativity and oblate surfaces. However, this modelling requires an accurate knowledge of the hot spot shape, its temperature profile, the stellar oblateness and the general-relativistic environment. All these ingredients are still uncertain, especially for millisecond pulsars, deviating significantly from a perfect sphere. Nevertheless, these observational hints tend to reliably constrain the hot spot location on the star and their area. In the present work, we start from a theoretical perspective and construct polar cap areas from the knowledge of the electromagnetic field in the force-free regime of a prolate or an oblate star starting from a theoretical point of view.

Computing light curves with general-relativistic effects has been done by \cite{cadeau_light_2007} and \cite{morsink_oblate_2007} who used an oblate Schwarzschild approximation to the exterior gravitational field. \cite{algendy_universality_2014} studied the impact of fast rotation on the gravitational field on the surface of a neutron star. They also gave analytical fits to the equatorial radius. Recently \cite{silva_surface_2021} constructed equilibria configurations using realistic neutron star equations of state. Such investigations give a good estimate of the expected stellar deformation due to centrifugal forces and helps to constrain the oblateness knowing the neutron star rotation period.
 
It is the purpose of this paper to compute force-free neutron star magnetospheres for a spheroidal star surrounded by an ideal pair plasma. In section~\ref{sec:simulations} we present the simulation set up, solving the time-dependent Maxwell equations using our fully pseudo-spectral code. In section~\ref{sec:results}, we plot magnetic field lines, spin down luminosities, polar cap shapes and current densities. Some discussion about fast rotating neutron stars is proposed in section~\ref{sec:discussion}. Conclusions and perspectives are drawn in section~\ref{sec:conclusion}.

\section{Time-dependent simulation setup}
\label{sec:simulations}

The present study relies on our previous work \citep{petri_spheroidal_2021}, adding an electric charge and current density in the force-free approximation. We succinctly remember the model before diving into the salient features of a force-free spheroidal magnetosphere.

The stellar interior is treated as a perfect conductor of spheroidal shape, being oblate or prolate. The time-dependent Maxwell equations are solved in a spheroidal coordinate system with coordinates~$(\rho,\psi,\phi)$ and oblateness or prolateness parameter~$a$. The coordinate system belongs to one of the eleven separable coordinate systems and forms an orthogonal basis. The electric and magnetic fields $\vec{E}$ and $\vec{B}$ are evolved in time according to
\begin{subequations}
	\begin{align}
	\label{eq:Maxwell1}
	\divg \vec{B} & = 0 \\
	\label{eq:Maxwell2}
	\rot \vec{E} & = - \frac{\partial \vec{B}}{\partial t} \\
	\label{eq:Maxwell3}
	\divg \vec{E} & = \frac{\rho_{\rm e}}{\varepsilon_0} \\
	\label{eq:Maxwell4}
	\rot \vec{B} & = \mu_0 \, \mathbf j + \frac{1}{c^2} \, \frac{\partial \vec{E}}{\partial t}  .
	\end{align}
\end{subequations}
$\varepsilon_0$ and $\mu_0$ are the electric permittivity and magnetic permeability respectively, $\rho_{\rm e}$ is the charge density and $\vec{j}$ the current density.
For an extensive and detailed discussion of the spheroidal coordinate systems, we refer to \cite{petri_spheroidal_2021} in order to avoid reproducing lengthy formulae here.

As usual in the relativistic force-free regime, the electric current density is deduced from the force-free prescription and reads
\begin{equation}
\label{eq:J_ideal}
\vec{j} = \rho_{\rm e} \, \frac{\vec{E}\wedge \vec{B}}{B^2} + \frac{\vec{B} \cdot \rot \vec{B} / \mu_0 - \varepsilon_0 \, \vec{E} \cdot \rot \vec{E}}{B^2} \, \vec{B} .
\end{equation}
The system of equation is therefore fully determined. 

The numerical setup for performing simulations is similar to the one used in \cite{petri_spheroidal_2021}. The gravitating fluid outer boundary is denoted by~$\rho_{\rm in} = R$ and should not to be confused with the spherical radius because we use spheroidal coordinates. The light-cylinder radius is $\rlight$ and the spheroidal inner surface of the computational domain set to $\rho_{\rm in} / \rlight=0.3$. The outer boundary of the computation domain is located at $\rho_{\rm out}/\rlight = 7$. We impose outgoing wave boundary conditions at the outer edge of the simulation domain and impose tangential electric field component as well as normal magnetic field component continuity on the stellar surface. A numerical convergence analysis showed that a grid resolution of $N_\rho \times N_\psi \times N_\varphi = 129\times32\times64$ was sufficient to reckon accurate results.

\section{Results}
\label{sec:results}

As physical relevant outputs of our simulations, we draw attention to the magnetic field structure, to the spin down luminosity, to the polar cap geometry and to the polar cap current density. These quantities are the baselines underlying a more detailed investigation of observational consequences like thermal X-ray emission, radio and gamma-ray pulse profiles. However computation of such emission processes are postponed to future works.

\subsection{Magnetic field lines}

In the force-free regime, the magnetic field lines are attached to the particles. Because outside the light-cylinder these particles can no more corotate with the star, field lines are opening up, the poloidal component tending asymptotically to radial lines wound up by rotation. This winding up is a general feature of any force-free magnetosphere reaching ultra-relativistic corotating speeds close to the light cylinder region and trying to enforce corotation beyond, irrespective of the exact inner boundary on the stellar surface. We verify this assertion by plotting some of these field lines for spherical, oblate and prolate stars.

\begin{figure*}
	\centering
	\includegraphics[width=\linewidth]{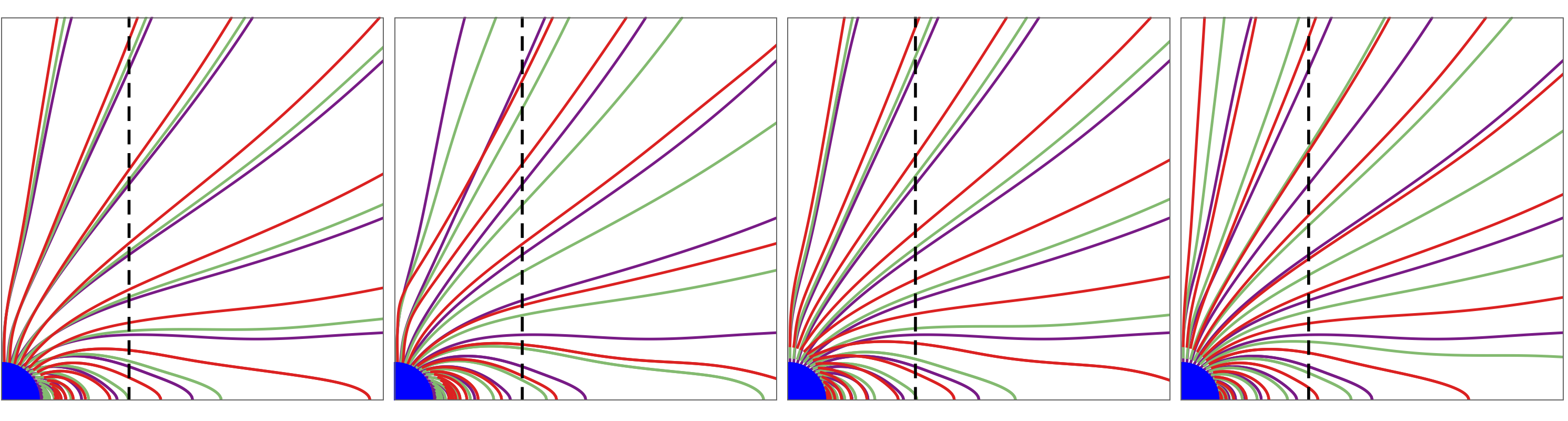} 
	\caption{Magnetic field lines in the meridional plane $xOz$ for an aligned spheroidal rotator with oblateness (first two columns) or prolateness (last two columns) parameter $a/R=\{0,0.5,1\}$ respectively in blue, green and red. The blue quarter disk in the bottom left depicts the spherical star. The vertical dashed black line represents the light-cylinder.}
	\label{fig:lignes_champ_B_xz_a0}
\end{figure*}

The asymptotic radial structure of poloidal field lines and the transition between closed and open field lines for an aligned rotator are shown quantitatively in Fig.~\ref{fig:lignes_champ_B_xz_a0} for an oblateness parameter $a/R=\{0,0.5,1\}$ in blue, green and red respectively. The blue quarter disk on the bottom left represents the spherical star. The light-cylinder is depicted as a vertical dashed black line. An oblate star inflates in the equatorial direction whereas a prolate star inflates along the rotation axis. Field lines of spheroidal stars do therefore not start right at the blue disk but at a larger distance.

\begin{figure*}
	\centering
	\includegraphics[width=\linewidth]{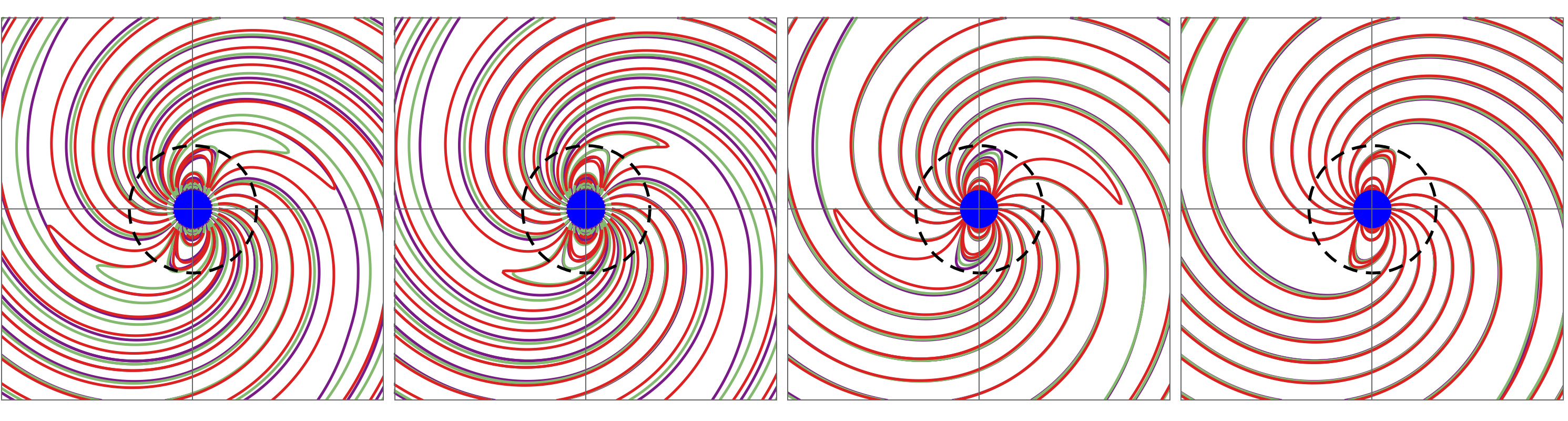}
	\caption{Magnetic field lines in the equatorial plane $xOy$ for a perpendicular spheroidal rotator with oblateness (first two columns) or prolateness (last two columns) parameter $a/R=\{0,0.5,1\}$ respectively in blue, green and red. The blue disk in the centre depicts the spherical star. The dashed black circle represents the light-cylinder.}
	\label{fig:lignes_champ_B_xy_a90}
\end{figure*}

Field lines in the equatorial plane for orthogonal rotators are shown in Fig.~\ref{fig:lignes_champ_B_xy_a90} with the same spheroidal parameter $a/R=\{0,0.5,1\}$ in blue, green and red respectively. The blue disk in the centre represents the spherical star. The two-armed spiral shape, typical of a rotating dipole, develops from the light-cylinder up to large distances. Its location in space is identical for all simulations.


\subsection{Spin down}

The spin down depends on the surface shape but also on the normalisation convention used to compute the luminosity. This important issue is discussed in depth by \cite{petri_spheroidal_2021}. In order to get rid of effects not directly related to the change in shape, we used a normalisation where the magnetic dipole tends asymptotically to the same magnetic moment value at infinity, irrespective of the oblateness or prolateness parameter~$a/R$.

\begin{figure}
	\centering
	\includegraphics[width=\linewidth]{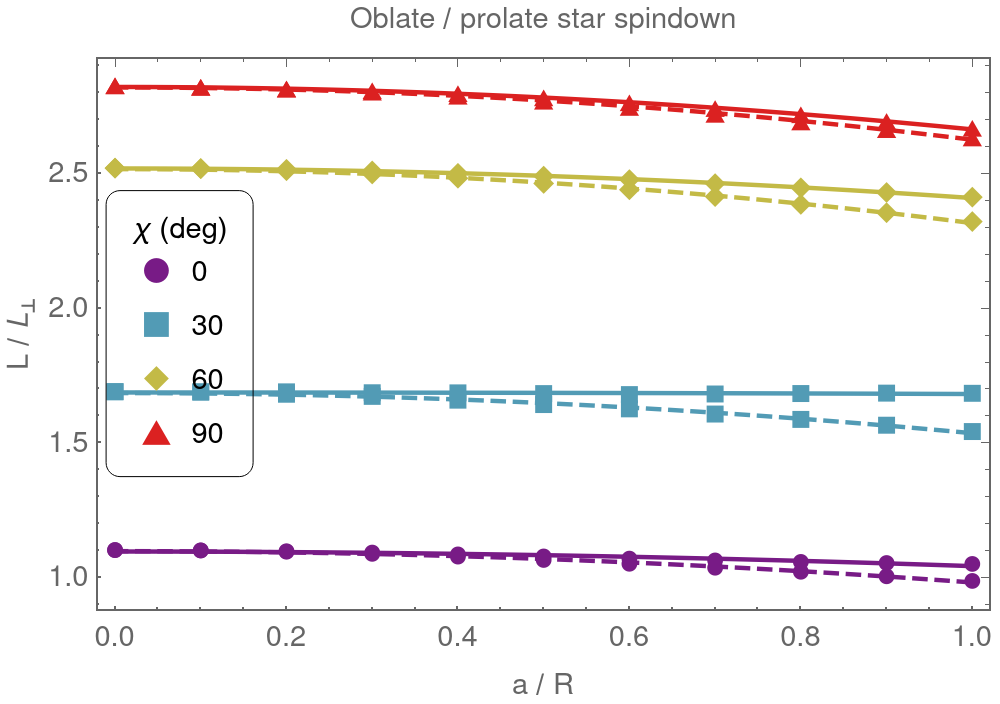} 
	\caption{Spin-down luminosity for oblate and prolate stars, respectively in solid and dashed lines, with single dipole stellar boundary conditions.}
	\label{fig:luminosite_ellipticite_ffe_112}
\end{figure}

\begin{figure}
	\centering
	\includegraphics[width=\linewidth]{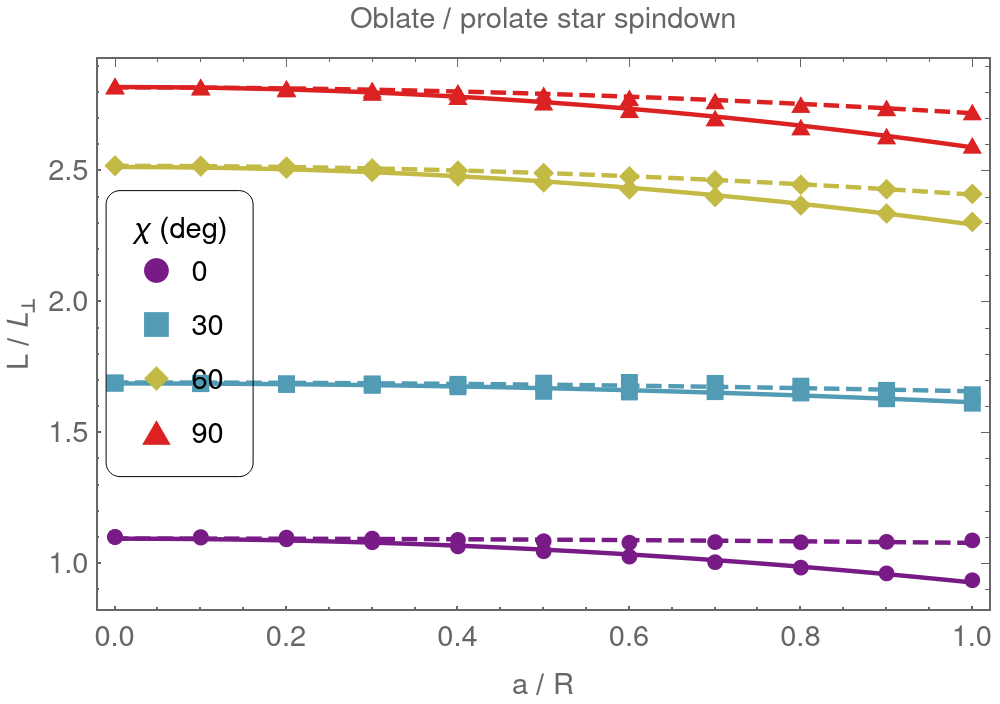}
	\caption{Spin-down luminosity for oblate and prolate stars, respectively in solid and dashed lines, with spherical dipole stellar boundary conditions.}
	\label{fig:luminosite_ellipticite_ffe_134}
\end{figure}

As for the vacuum case, the spin down decreases with increasing parameter~$a/R$, as seen in Fig.~\ref{fig:luminosite_ellipticite_ffe_112} for single dipole field on the surface and in Fig.~\ref{fig:luminosite_ellipticite_ffe_134} for spherical dipole imposed on the surface.

All simulation runs are summarized by a spin down fitted formally with 
\begin{equation}\label{eq:Luminosite_fit}
 L_{\rm FFE}^{\rm spheroid} \approx L_\perp \, (\ell_1 + \ell_2 \, \sin^2 \rchi) ,
\end{equation}
where the vacuum luminosity for a perpendicular rotator is
\begin{equation}\label{eq:Lperp}
L_\perp = \frac{8\,\pi}{3\,\mu_0\,c^3}\,\Omega^4\,B^2\,R^6 ,
\end{equation}
where $B$ is the magnetic field strength at the equator of a spherical star, $\Omega$ its rotation rate and $R$ its radius.
The coefficients~$\ell_i$ for $i=\{1,2\}$ depend on the ellipticity according to
\begin{equation}\label{eq:Fit}
\ell_i = \alpha_i - \beta_i \, \left( \frac{a}{R} \right)^2 .
\end{equation}
The fitted values are reported in table~\ref{tab:fit}.
\begin{table}
	\centering
	\begin{tabular}{|c|cccc|}
		\hline
		Model & $\alpha_1$ & $\beta_1$ & $\alpha_2$ &  $\beta_2$ \\
		\hline
		oblate  & 1.17 & 0.0212 & 1.71 & 0.116 \\
		prolate & 1.14 & 0.120 & 1.71 & 0.0975 \\
		oblate spherical  & 1.17 & 0.123 & 1.71 & 0.0830 \\
		prolate spherical & 1.18 & 0.0172 & 1.71 & 0.0932 \\
		\hline
	\end{tabular}
	\caption{Fitted coefficients $\alpha_i$ and $\beta_i$ as given by eq.~\eqref{eq:Fit}.\label{tab:fit}}
\end{table}
The spheroidal shape always decreases the spin down luminosity with respect to the spherical star. This effect is most pronounced for prolate stars.

\subsection{Polar cap shape}

\begin{figure}
	\centering
	\includegraphics[width=\linewidth]{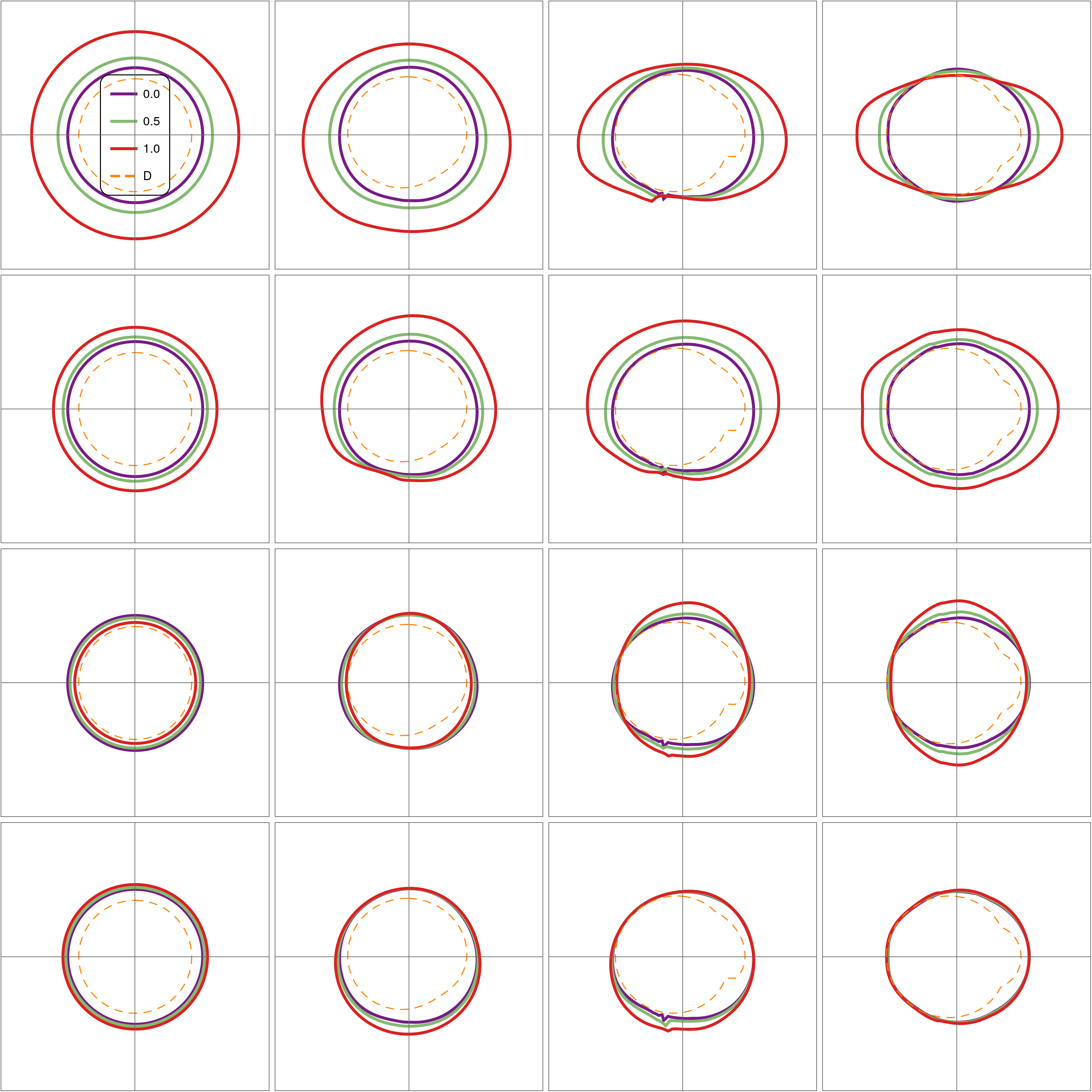}
	\caption{Polar cap shape for oblate and prolate stars with oblateness parameter $a/R=\{0,0.5,1\}$ respectively in blue, green and red. The orange dashed line shows the reference solution for the Deutsch field as a check. The obliquity from the left column to the right column is $\rchi=\{0\degr, 30\degr, 60\degr, 90\degr\}$. First row for an oblate star with one multipole of order $\ell=1$, second row for a spherical dipole magnetic field at the surface, third row for a prolate star with one multipole of order $\ell=1$ and fourth row for a spherical dipole magnetic field at the surface.}
	\label{fig:forme_calotte}
\end{figure}

Polar caps are important observables. Indeed, the shape of their rims can be indirectly probed by their thermal X-ray pulsation as detected by for instance the Neutron Star Interior Composition Explorer (NICER) \citep{riley_nicer_2019,bogdanov_constraining_2019}. As a starting point for computing thermal X-ray emission, we present in this paper realistic polar cap surfaces relying on our force-free simulations of spheroidal stars. Fig.~\ref{fig:forme_calotte} compiles the geometry of the polar cap edges for various neutron star surface deformations, being either oblate or prolate, different inclination angles~$\rchi=\{0\degr, 30\degr, 60\degr, 90\degr\}$ and several normalized spheroidal parameter~$a/R=\{0,0.5,1\}$. For reference, the vacuum Deutsch solution is also shown in dashed orange lines.

Compared to polar caps deduced from vacuum, in force-free magnetospheres, the cap rims lie always outside the equivalent Deutsch case. For the aligned rotator, the size and area of the cap, being perfectly circular due to the axisymmetry, increases with oblateness or prolateness, except for the third raw, as seen in the left column of Fig.~\ref{fig:forme_calotte}. For large obliquities, the polar cap surface area increases also with the spheroidal parameter~$a$. Nevertheless, for prolate stars, the deformation of the polar cap remains weak because along the equator, the stellar shape remains almost spherical, tending to the spherical force-free results. For oblate stars, the situation is opposite because the maximum deformation arises around the equator and the polar cap area increases significantly with oblateness.
Contrary to the spin down comparison performed in the previous section where some fiducial spherical model had to be chosen, the polar cap rims rely solely on geometrical effects of magnetic field lines. Therefore the polar surfaces discussed in this section are insensitive to the particular normalisation used for the magnetic dipole. The results are robust and can applied to any magnetic field intensity.

\subsection{Polar cap current}

Just like the previous quantities, the current produced within the magnetosphere is significantly altered by the spheroidal shape of the stellar surface. The local current density expelled and returning to the polar caps represents offers an important view of the electrodynamics impacted by the spheroidal star. This current density is best reckoned in the star corotating frame, removing the convective current produced by the electric drift motion of the charge-separated plasma component. This current is not directly related to the conduction current which is the primordial current to plot in force-free electrodynamics. \cite{endean_lorentz_1974} presented an interesting derivation of the relativistic force-free equation involving only the magnetic field, therefore generalizing the usual non relativistic equation 
\begin{equation}\label{rq:FFEnonrel}
( \vec{\nabla} \wedge \vec{B} ) \wedge \vec{B} / \mu_0 = \vec{j} \wedge \vec{B} = \vec{0}
\end{equation}
to a relativistic version given by
\begin{equation}\label{eq:FFErel}
( \vec{\nabla} \wedge \vec{B}^* ) \wedge \vec{B} / \mu_0 = \vec{0}
\end{equation}
where he introduced a new vector $\vec{B}^*$ defined by
\begin{equation}\label{eq:FFE_Bcor}
\vec{B}^* = \left( \left(1 - \frac{x^2+y^2}{\rlight^2}\right) \, B^\rho, \left(1 - \frac{x^2+y^2}{\rlight^2}\right) \, B^\psi, B^\phi \right).
\end{equation}
A similar expression was given by \cite{mestel_force-free_1973} at the same time and also explained by a simple Lorentz transformation. Therefore according to Eq.~\eqref{eq:FFErel}, the term 
\begin{equation}\label{eq:jconduction}
\mu_0 \, \vec{j}_c = \vec{\nabla} \wedge \vec{B}^*
\end{equation}
is directed along $\vec{B}$ and can be interpreted as the conduction current density~$\vec{j}_c$ in the corotating frame, see also \cite{bai_modeling_2010} for another derivation of this interpretation. For numerical purposes, we normalize the current density to a fiducial value related to the corotating current density at the pole of an aligned rotator and given by
\begin{equation}\label{key}
j_{\rm pol} = 2 \, \varepsilon_0 \, \Omega \, B \, c . 
\end{equation}

\begin{figure}
	\centering
	\includegraphics[width=0.9\linewidth]{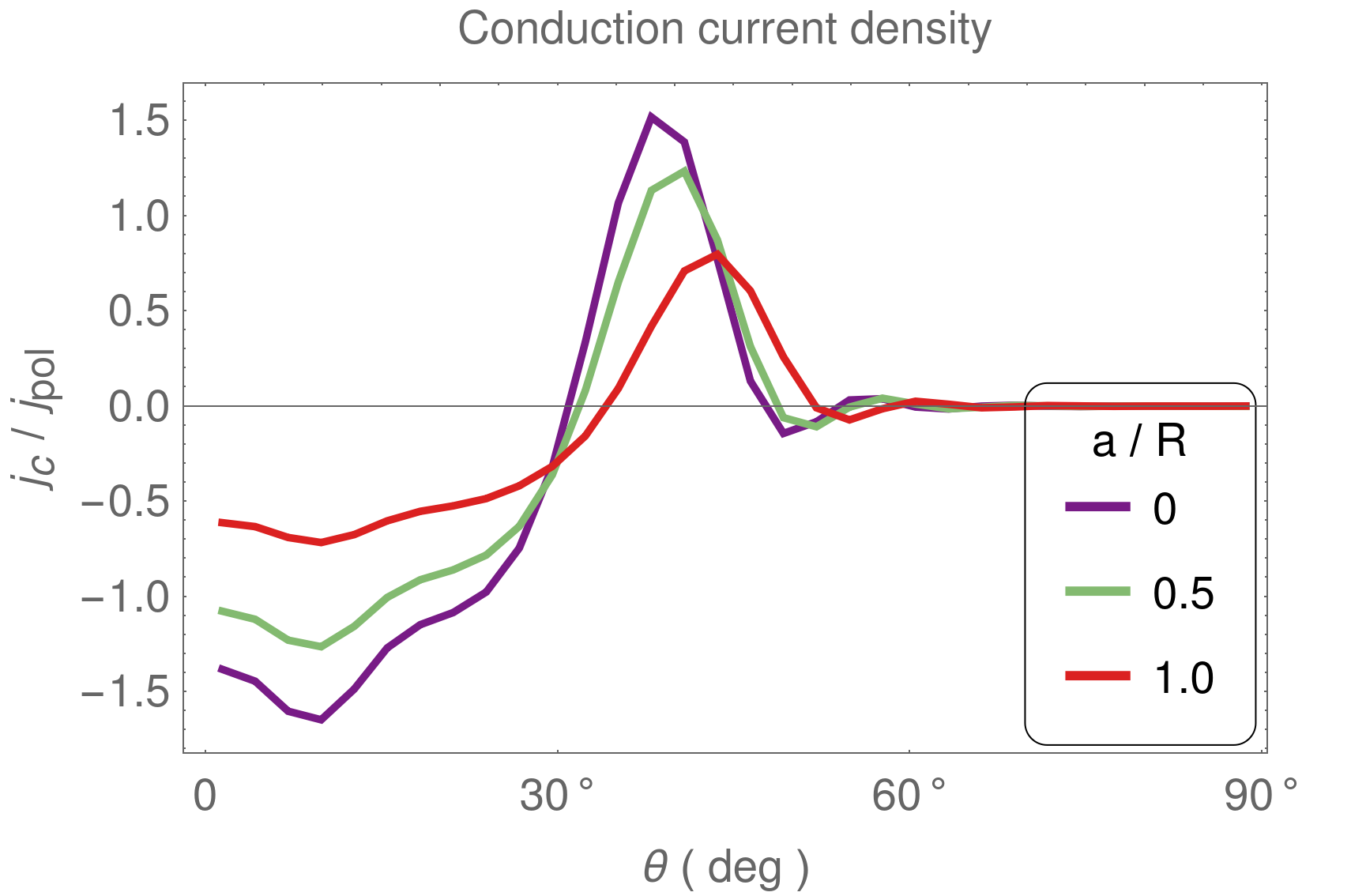}
	\caption{Conduction current density on the surface of an aligned oblate star with oblateness parameter $a/R=\{0,0.5,1\}$.}
	\label{fig:jconduction_l11}
\end{figure}

Following the definition given by Eq.~\eqref{eq:jconduction} fig.~\ref{fig:jconduction_l11} shows the conduction current density on the surface of an aligned oblate star with oblateness parameter $a/R=\{0,0.5,1\}$ for one pole with $\theta \in [0\degr, 90\degr]$. The other pole is not shown because of the north-south antisymmetry. The conduction current $\vec{j}_c$ is actually symmetric with respect to the equator located at a colatitude $\theta=90\degr$. The oblateness decreases the current density by up to a factor two for $a/R=1$. Nevertheless the polar cap area also increases with oblateness, partially compensating for the current density decrease. The current flowing in the vicinity of the polar cap therefore depends crucially on the oblateness. Combining the variation in the area, we expect the associated thermal X-ray emission to be drastically influenced by the stellar surface geometry in addition to the polar cap area and shape.

\begin{figure}
	\centering
	\includegraphics[width=0.9\linewidth]{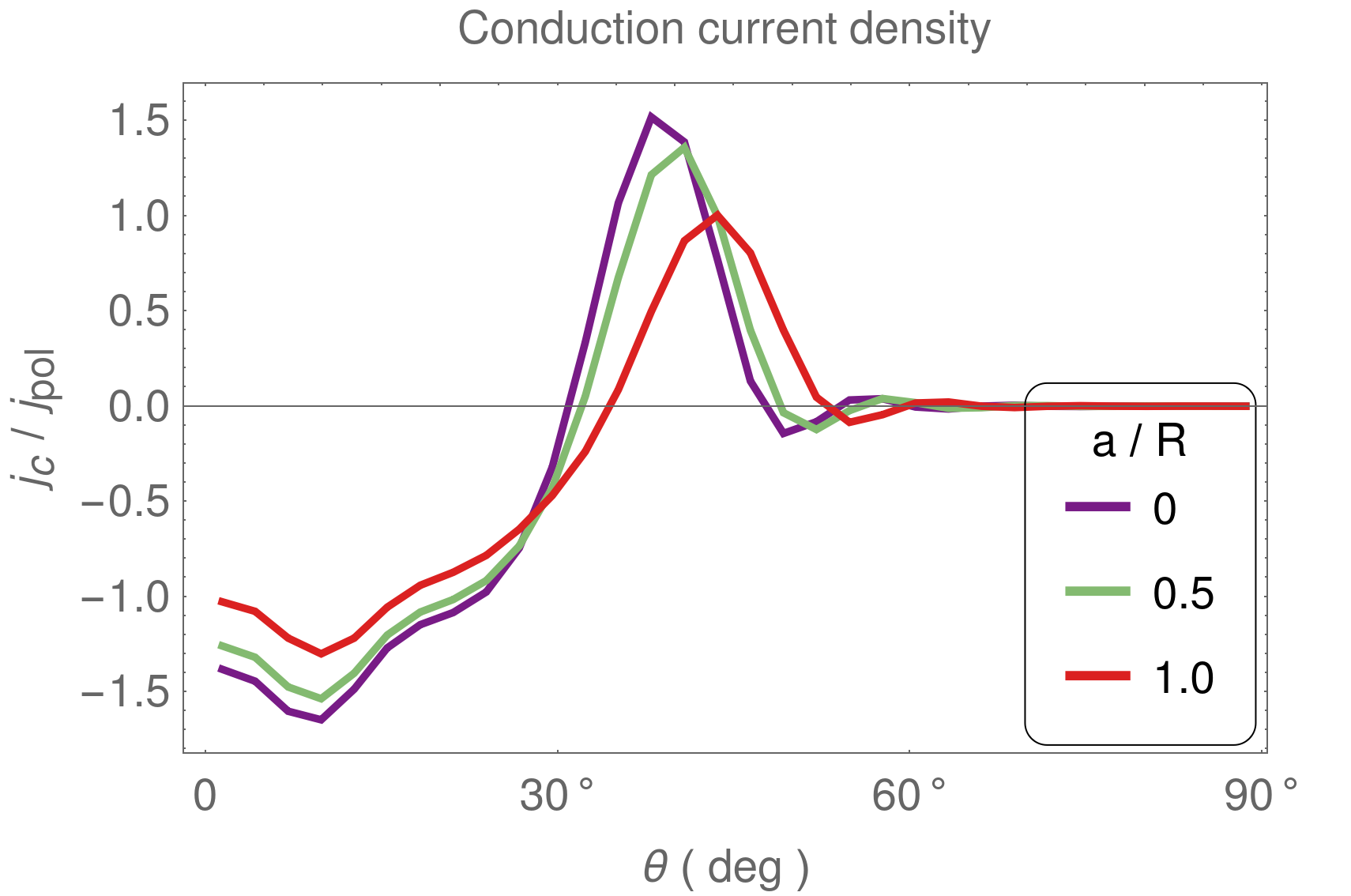}
	\caption{Conduction current density on the surface of an aligned prolate star with oblateness parameter $a/R=\{0,0.5,1\}$.}
	\label{fig:jconduction_l12}
\end{figure}

Fig.~\ref{fig:jconduction_l12} shows the same conduction current density on the surface of an aligned prolate star with oblateness parameter $a/R=\{0,0.5,1\}$. Here also, the current density decreases but to a lesser extend compared to the oblate case. The impact on the thermal X-ray emission is consequently less affected than in the previous oblate case.

\begin{figure}
	\centering
	\includegraphics[width=0.9\linewidth]{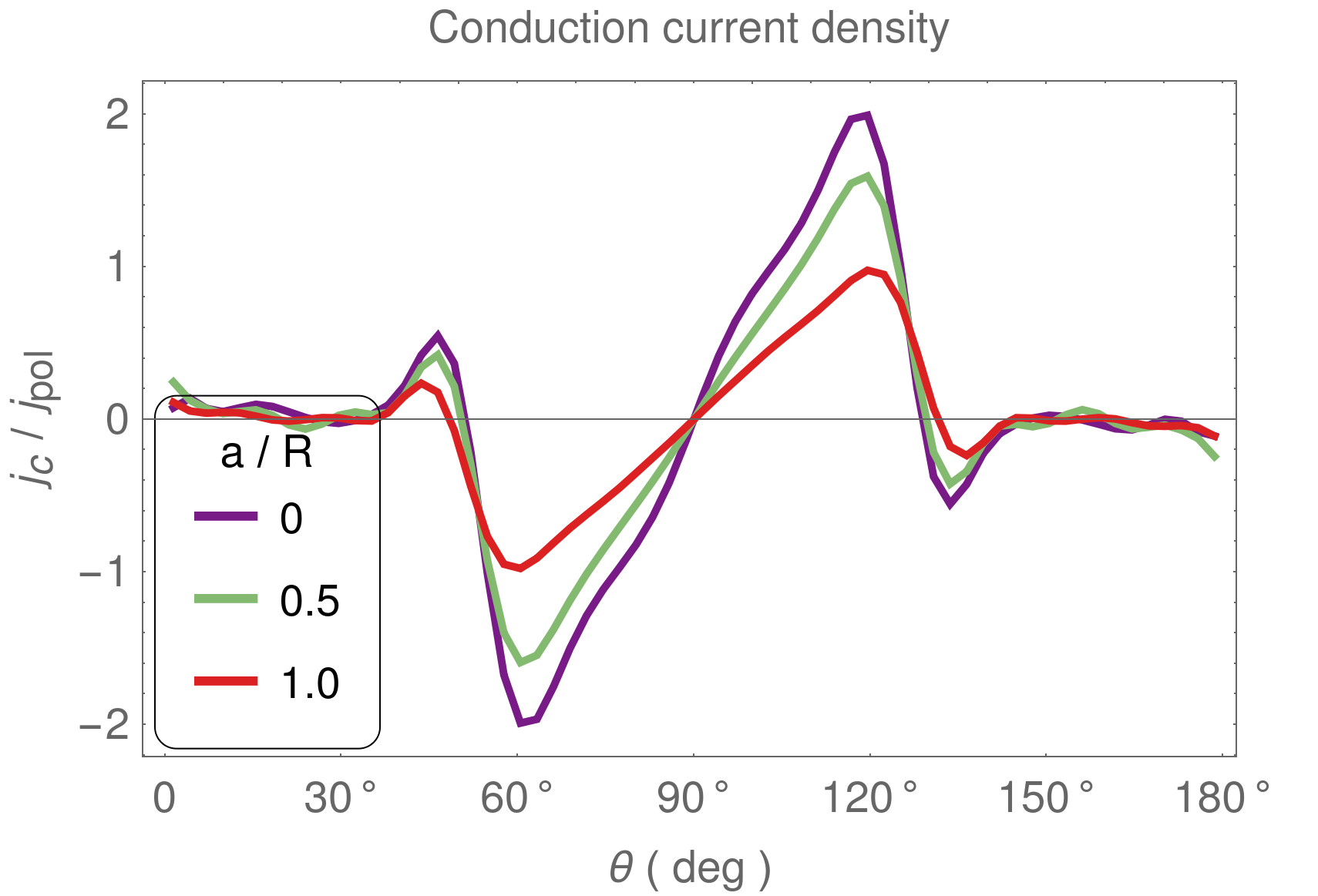}
	\caption{Conduction current density on the surface of an orthogonal oblate star with oblateness parameter $a/R=\{0,0.5,1\}$.}
	\label{fig:jconduction_l11_a90}
\end{figure}

\begin{figure}
	\centering
	\includegraphics[width=0.9\linewidth]{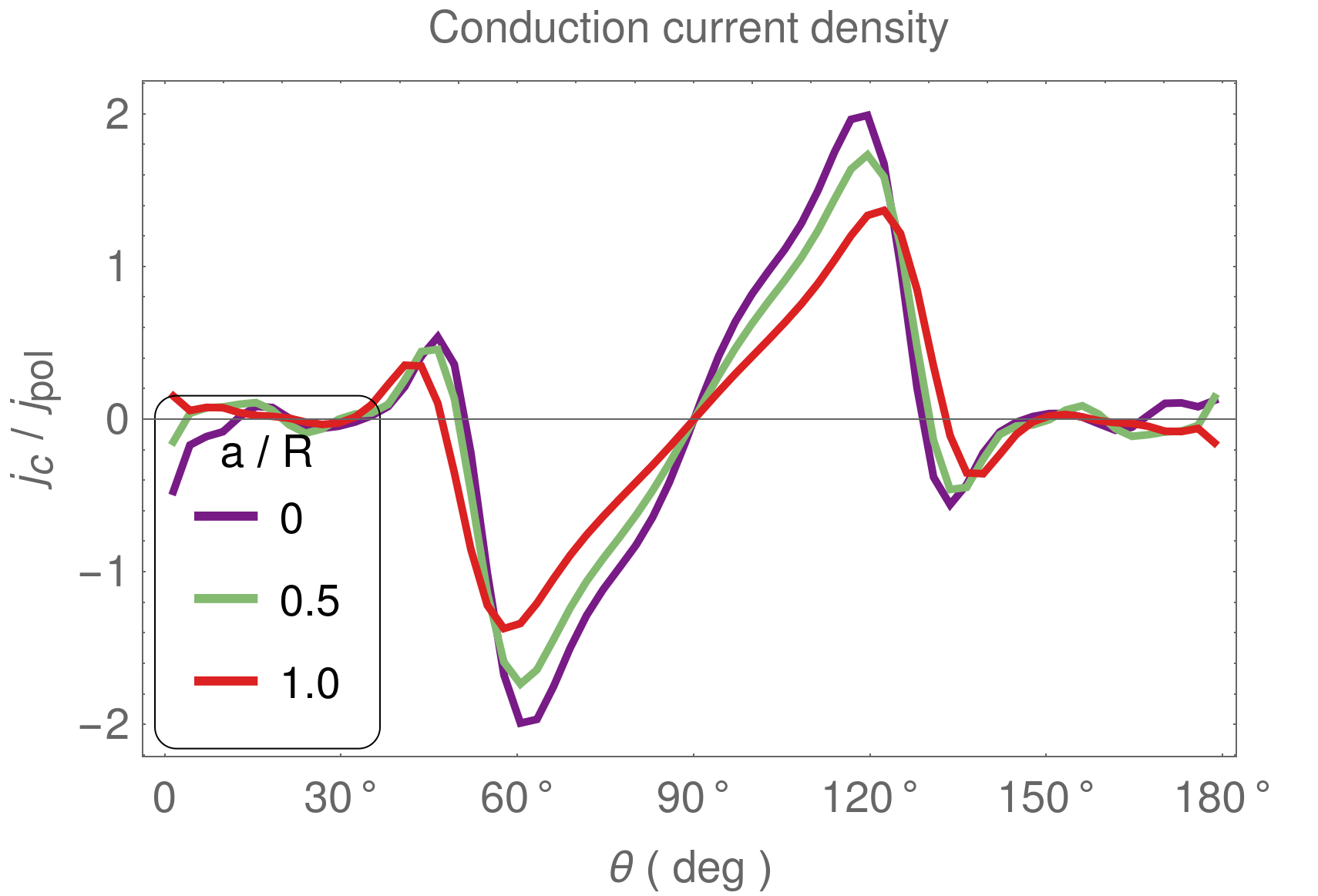}	
	\caption{Conduction current density on the surface of an orthogonal prolate star with prolateness parameter $a/R=\{0,0.5,1\}$.}
	\label{fig:jconduction_l12_a90}
\end{figure}

To be exhaustive, we also computed the current density for orthogonal rotators in oblate and prolate geometries. Fig.~\ref{fig:jconduction_l11_a90} shows a cross section, taken around the centre of the polar cap, of the conduction current in oblate stars with oblateness parameter $a/R=\{0,0.5,1\}$. The conduction current $\vec{j}_c$ is actually skew-symmetric with respect to the equator located at a colatitude $\theta=90\degr$.
Eventually, fig.~\ref{fig:jconduction_l12_a90} shows the same cross section for a prolate star with prolateness parameter $a/R=\{0,0.5,1\}$. We always notice a monotonic decrease in the peak current density with increasing spheroidal parameter~$a$.

\section{Discussion}
\label{sec:discussion}

Nowadays, thermal X-ray pulsation emanating from the surface of neutron stars is used to probe the location and shape of the underlying hot spots. Recent observations by NICER clearly demonstrated the strength of such approaches to constrain the mass, the equatorial radius \citep{riley_nicer_2019} and the equation of state \citep{miller_psr_2019} of these compact objects. However, these investigations rely on observational fitting that are not always quantitatively and accurately linked to neutron star magnetosphere models. For instance, we are not aware of any oblate force-free magnetosphere computation showing the impact of oblateness on spin down efficiency and polar cap perturbations induced by the deviation form a spherical star.

Our results unambiguously predicts the importance of carefully estimating the oblateness in order to fit hot spot X-ray light curves. Irrespective of the normalization employed for the magnetic field, we always found an increase in the hot spot surface area compared to a spherical star. The current flowing back to the star, hitting and heating the surface is therefore also modified by the oblateness or prolateness. As a consequence, the surface brightness temperature and the luminosity of these hot spots are influenced by the stellar boundary, not only because of geometrical effects but also because of electrodynamics effects related to the change in current and charge density existing inside the magnetosphere.

Another important ingredient not included in the present study is general relativity. Because of the large compactness of neutron stars, curved space time also significantly modifies the electrodynamics of its magnetosphere, especially in the vicinity of its surface. Unfortunately this requires solutions to Einstein equations for an oblate gravitating fluid and simple analytical formulas for the exterior space time are difficult to find with exception of \cite{erez_gravitational_1959}, corrected by \cite{young_exact_1969} and \cite{doroshkevich_gravitational_1965}. A very general solution for the exterior of the star has been proposed by \cite{quevedo_general_1989}. The multipole components are however left as free parameters not straightforwardly connected to the stellar oblateness. Nevertheless, a simple and good approximation is obtained by the so-called oblate+Schwarzschild model where the oblate stellar surface is embedded into the Schwarzschild metric. It is simple but furnishes a good impression on these additional effects even if it does not include frame-dragging.
Last but not least, it was shown that magnetic multipole components \citep{bilous_nicer_2019} can play a central role in understanding millisecond pulsar light-curves \citep{kalapotharakos_multipolar_2021}. 

The oblateness parameter $a$ can be deduced from numerical models of neutron star interiors. According to table~I of \cite{silva_surface_2021}, using realistic equations of state for the construction of stars in stationary equilibrium for fast rotating neutron stars with frequency about 600~Hz, the ratio between polar radius~$R_p$ and equatorial radius~$R_e$ is about $r = R_p/R_e \approx0.9$, implying an oblateness parameter of $a = \sqrt{r^{-2}-1} \approx 0.48$. The deformation of the stellar surface is therefore substantial, leading to a slight increase in the polar cap surface area. 

The impact on the spin down luminosity depends on the internal structure of the magnetic field compared to the equivalent spherical star. This issue is however not solved. For some neutron stars, a birth period around 1~ms is expected, even for strongly magnetized ones, coined millisecond magnetars and suspected to be the central engine of hypernovae or superluminous supernovae. For such high periods, the aspect ratio is about $r \approx 0.7$, implying an oblateness of $a \approx 1$. If we assume an ideal plasma compression due to the slowing down of the star, the magnetic flux is approximately conserved and the magnetic field strength at the equator increases by a factor $1/r \approx 1.42$. The associated spin down augments by a factor $1/r^2 \approx 2.0$. Therefore at least two mechanisms contribute to a modification of the neutron star luminosity, first the change in the stellar surface and second the variation in the surface magnetic field strength. We believe that these effects must be taken into account for the evolution of the early stage of a neutron star relaxing to a spherical shape as it slows down.

\section{Conclusions}
\label{sec:conclusion}

Although neutron stars are subject to enormous gravitational binding energy, millisecond rotation periods can lead to significant centrifugal forces deforming their surface into an oblate shape or even into a prolate shape due to magnetic compression. We computed force-free magnetospheres for spheroidal neutron stars and showed the repercussion onto magnetic topology, spin down efficiency, polar cap geometry and current density. We observed significant changes in the polar cap rims for reasonable spheroidal parameters. Compared to a spherical star, the impact on spin down luminosity remains small with our normalisation convention. However the magnetic field geometry repercussing on the polar cap shape is significant irrespective of the magnetic field strength. Such shapes can indirectly be probed via thermal X-ray emission from hot spots as recently demonstrated by the NICER collaboration.

From our study, we can compute such X-ray light-curves, knowing the size and shape of the hot spot. Nevertheless, in order to better stick to realistic neutron stars, we need to take into account the compactness of the star and therefore are obliged to include some general-relativistic effects using for instance an oblate Schwarzschild metric or more accurate models of spheroidal neutron star gravitational fields. These investigations are the natural steps that we pursue in forthcoming studies.

\begin{acknowledgements}
I am grateful to the referee for helpful comments and suggestions.
This work has been supported by the CEFIPRA grant IFC/F5904-B/2018 and ANR-20-CE31-0010.
\end{acknowledgements}

\bibliographystyle{/home/petri/publications/bibtex/aa}
\bibliography{/home/petri/zotero/Ma_bibliotheque}

\end{document}